\documentclass{article}
\usepackage{graphicx}
\usepackage{cite}
\begin{document}

\title{\bf $\lambda\phi^{4}$ Kink and sine-Gordon Soliton in the GUP Framework}
\author{M. Asghari and P. Pedram\thanks{Email: p.pedram@srbiau.ac.ir},
\\  {\small Department of Physics, Tehran Science and Research Branch, Islamic Azad University, Tehran, Iran}}

\maketitle \baselineskip 24pt

\begin{abstract}
We consider $\lambda\phi^{4}$ kink and sine-Gordon soliton in the
presence of a minimal length uncertainty proportional to the Planck
length. The modified Hamiltonian contains an extra term proportional
to $p^4$ and the generalized Schr\"odinger equation is expressed as
a forth-order differential equation in quasiposition space. We
obtain the modified energy spectrum for the discrete states and
compare our results with 1-loop resummed and Hartree approximations
for the quantum fluctuations. We finally find some lower bounds for
the deformations parameter so that the effects of the minimal length
have the dominant role.
\end{abstract}



\section{Introduction}
At high energy limit the Heisenberg algebra will be modified by
adding small corrections to the canonical commutation relation in
the form of the Generalized Uncertainty Principle (GUP)
\cite{felder}. From these modifications a short distance structure
is obtained that characterized by a finite minimal uncertainty
$(\Delta X)_{min}$ in the position measurement. String theory, loop
quantum gravity, noncommutative geometry, and black-hole physics all
suggest the existence of such a minimal measurable length of the
order of the Planck length $\ell
_{\mathrm{P}}=\sqrt{\frac{G\hbar}{c^{3}}}\approx 10^{-35}$m. The
presence of this minimal observable length modifies all the
Hamiltonians and many papers have appeared in literature to address
the effects of GUP on various physical systems
\cite{1,2,3,4,5,6,7,8,9,10,11,12,13,14,15,16,17,18,19,20,21,22,23}.

A linear and dispersionless relativistic wave equation holds two
features, namely (i) keeping the shape and velocity of a single
packet and (ii) keeping the asymptotic shape and velocity of several
packets even after collisions \cite{24}. However, there are much
more complicated wave equations in many branches of physics that
contain nonlinear terms, dispersive terms, and several coupled wave
fields. Solitary waves are certain special solutions of nonlinear
wave equations that look like pulses of energy traveling without
dissipation and with uniform velocity. The solitary waves whose
energy density profiles are asymptotically restored to their
original shapes and velocities are known as solitons. However, Kink
solutions of $\lambda\phi^{4}$ theory are solitary waves but not
solitons. At the classical level they resemble extended particles,
i.e., localized and finite-energy objects. On the other hand, the
solutions of the sin-Gordon system are solitary waves and also
soliton. This model has been used in the study of a wide range of
phenomena such as propagation of crystal dislocations, of magnetic
flux in Josephson lines and two-dimensional models of elementary
particles \cite{24-1,24-2}.

Here, we study $\lambda\phi^{4}$ kink and sin-Gordon soliton in the
GUP framework given by the following generalized uncertainty
relation \cite{12}
\begin{equation}\label{0}
 \Delta X\Delta P\geq \frac{\hbar}{2}(1+\beta(\Delta P)^{2}+\gamma),
\end{equation}
where $\beta$ is the GUP parameter and $\gamma=\beta\langle
P\rangle^{2}$. This inequality implies the existence of a minimum
observable length proportional to the square root of the deformation
parameter, i.e., $(\Delta X)_{min}=\hbar \sqrt{\beta}$. We obtain
the effects of this minimal length on kink and soliton energy
spectrums and compare our results with the ones obtained in 1-loop
resummed and Hartree approximations \cite{27,28}. This suggests some
lower bounds on the GUP parameter for each case.

This paper is organized as follows: In section 2, we state the
generalized  uncertainty principle and its first order
representation in quasiposition space. In section 3, we find
$\lambda\phi^{4}$ kink modified energy spectrum using the first
order perturbation theory and compare our results with 1-loop
resummed and Hartree approximations. In section 4, we outline the
energy correction in Hartree approximation for the sine-Gordon
soliton and obtain the GUP correction to its discrete energy level.
Finally, we present our conclusions in section 5.

\section{The Generalized Uncertainty Principle}
Let us consider the following deformed commutation relation
\begin{equation}\label{1}
\left [X,P\right]=i\hbar\left(1+\beta P^{2}\right),
\end{equation}
which leads to the generalized uncertainty relation (\ref{0}) and
$\beta=\beta_{0}/(M_{\mathrm{P}}c)^{2}$. Here $\beta_0$ is of the
order of unity and $M_{\mathrm{P}}$ is the Planck mass. We can
define \cite{19}
\begin{eqnarray}\label{3}
    X&=&x,\\
\label{4}
     P&=&\frac{\tan ( \sqrt{\beta} \, p )  }{\sqrt{\beta}},
\end{eqnarray}
to exactly satisfy the above modified uncertainty principle, where
$x$ and $p$ obey the canonical commutation relations, i.e.,
$\left[x,p\right]=i\hbar$. Moreover, $p$ can be interpreted as the
momentum operator at low energies $p=-i\hbar\frac{\partial}{\partial
x}$ and $P$ as the momentum operator at high energies. The general
form of the Hamiltonian
\begin{equation}\label{5}
   H=P^{2}+V(X)=\frac{\tan^2 ( \sqrt{\beta} \, p )  }{\beta}+V(x),
\end{equation}
to first order of the GUP parameter can be written as
\begin{equation}\label{6}
   H=H_{0}+ H_{1},
\end{equation}
where $H_{0}=p^{2}+V(x)$ and $H_{1}=\frac{2}{3}\beta p^{4}.$ Using
the expression for the Hamiltonian, the generalized form of the
Schr\"odinger equation in quasiposition space is given by
($\hbar=1$)
\begin{equation}\label{7}
-\frac{\partial^{2}\psi(x)}{\partial x^{2}}+\frac{2\beta}{3}\frac{\partial^{4}\psi(x)}{\partial x^{4}}+V(x)\psi(x)=E\psi(x),
\end{equation}
which has an extra term in comparison to ordinary Schr\"odinger
equation because of the deformed commutation relation (\ref{1}).
Since this equation is a 4th-order differential equation, it is not
an easy task in general to solve it in quasiposition space. Hence,
we implement the perturbation method in order to obtain the
solutions.

\section{$\lambda\phi^{4}$ kink in the GUP framework}
The $\lambda\phi^{4}$ theory with the Lagrangian density
\begin{equation}\label{8}
   {\cal L}(x,t)=\frac{1}{2}\left(\partial_{\mu}\phi\right)\left(\partial^{\mu}\phi\right)-V(\phi),
\end{equation}
where
$V(\phi)=(\lambda/4)\left(\phi^{2}-\frac{m^{2}}{\lambda}\right)^{2}$
leads to the following equation of motion
\begin{equation}\label{8/1}
    \partial_{\mu}\partial^{\mu}\phi+m^{2}\phi-\lambda \phi^{3}=0.
\end{equation}
Here, we focus on one of the solutions of the static equation,
namely, the classical kink solution $\phi_{\mathrm{kink}}(x)=\pm
\left(m/\sqrt{\lambda}\right)
\tanh\left[\left(m/\sqrt{2}\right)x\right]$, which interpolates
between the two degenerate vacuum states $\phi=\pm m/\sqrt{\lambda}$
\cite{24}. The fluctuation equation for $\eta(x)$ around this
solution is \cite{26}
\begin{equation}\label{9}
\left[-\frac{\partial^{2}}{\partial
x^{2}}+m^{2}\left(3\tanh^{2}\left(\frac{mx}{\sqrt{2}}\right)-1\right)\right]\eta_{n}(x)=\omega_{n}^{2}\eta_{n}(x),
\end{equation}
which besides its zero mode
\begin{eqnarray}\label{10}
\eta_{0}(x)=\sqrt{\frac{3m}{4\sqrt{2}}}\,\mbox{sech}^{2}\left(\frac{m x}{\sqrt{2}}\right),\hspace{3.75cm}\omega_{0}^{2}=0,
\end{eqnarray}
contains another bound state
\begin{eqnarray}\label{11}
\eta_{1}(x)=\sqrt{\frac{3m}{2\sqrt{2}}}\sinh\left(\frac{m
x}{\sqrt{2}}\right)\mbox{sech}^{2}\left(\frac{m x}{\sqrt{2}}\right),\hspace{2cm}\omega_{1}^{2}=\frac{3}{2}
m^{2}.
\end{eqnarray}
In quantum theory we construct a set of approximate harmonic
oscillator states around the point $\phi_{\mathrm{kink}}(x)$ in
field space and we expect the energies of these states to be the
kink particle associate with the lowest energy level. Using
Eqs.~(\ref{6}) and (\ref{9}) we get the expression for the
Hamiltonian of the kink system in the presence of GUP as
\begin{equation}\label{11/1}
H=p^{2}+\frac{2}{3}\beta p^{4}+m^{2}\left(3\tanh^{2}\left(\frac{m
x}{\sqrt{2}}\right)-1\right),
\end{equation}
where $H_{0}=p^{2}+m^{2}\left(3\tanh^{2}\left(\frac{m
x}{\sqrt{2}}\right)-1\right)$ is the Hamiltonian of the unperturbed
system. Now, we apply the perturbation theory to calculate the kink
energy corrections to ${\cal O}(\beta)$ as
\begin{eqnarray}\label{12}
\Delta\omega_{0}^{2}=\frac{2}{3}\beta\langle\eta_0|p^{4}|\eta_0\rangle=\frac{8}{21}\beta
m^{4},
\end{eqnarray}
and
\begin{eqnarray}\label{13}
\Delta\omega_{1}^{2}=\frac{2}{3}\beta\langle\eta_1|p^{4}|\eta_1\rangle=\frac{31}{42}\beta
m^{4}.
\end{eqnarray}
On the other hand, based on Ref.~\cite{27}, by knowing the
correlation function $G(x)$ of the following equation
\begin{equation}\label{14}
\left[-\frac{\partial^{2}}{\partial
x^{2}}+m^{2}\left(3\tanh^{2}\left(\frac{mx}{\sqrt{2}}\right)-1\right)+3\lambda
G(x)\right]\tilde{\eta}_{k}(x)=\tilde{\omega}_{k}^{2}\tilde{\eta}_{k}(x),
\end{equation}
the modified kink energy eigenvalues can be obtained in Hartree and
1-loop resummed approximations. The numerical results for the
eigenvalues in 1-loop resummed approximation are
$\tilde{\omega}_{0}^{2}=0.5 \lambda$ and
$\tilde{\omega}_{1}^{2}=1.71 \lambda$ \cite{27}, which lead to the
energy shifts
\begin{equation}\label{14/1}
 \Delta \tilde\omega_{0}^{2}=0.5 \lambda,\hspace{3cm} \Delta \tilde\omega_{1}^{2}=0.21
 \lambda.
\end{equation}
Also in Hartree approximation we have $\tilde{\omega}_{0}^{2}=0.33
\lambda$ and $\tilde{\omega}_{1}^{2}=1.29\lambda$  \cite{27}, so we
find
\begin{equation}\label{14/2}
    \Delta \tilde\omega_{0}^{2}=0.33\lambda,\hspace{3cm}\Delta \tilde\omega_{1}^{2}=-0.21 \lambda.
\end{equation}
Therefore, the comparison between the above results and ones
obtained in the GUP scenario shows that for
\begin{equation}
    \beta>1.32\frac{\lambda}{m^{4}},
\end{equation}
the effects of the minimal length or the discreetness of space are
more important than the effects that come from considering the
quantum fluctuations.

\section{sine-Gordon soliton in the GUP framework}
The sine-Gordon system is defined by a single scalar field
$\phi(x,t)$ in (1+1) dimensions governed by the Lagrangian density
\cite{24}
\begin{equation}
{\cal
L}(x,t)=\frac{1}{2}\left(\partial_{\mu}\phi\right)\left(\partial^{\mu}\phi\right)+\frac{m^{2}}{g^{2}}\left(\cos
g \phi-1\right),
\end{equation}
which gives rise to the following sine-Gordon equation
\begin{equation}
    \partial_{\mu}\partial^{\mu}\phi+\frac{m^{2}}{g}\sin\left(g\phi\right)=0.
\end{equation}
One of the static localized solutions, namely,
\begin{equation}
   \phi_{sol}=\frac{4}{g}\arctan\left(e^{mx}\right),
\end{equation}
is called  soliton. The equation for the fluctuations around this
solution is expressed as the following Schr\"odinger-like equation
\cite{25}
\begin{equation}\label{15/1}
\left[-\frac{\partial^{2}}{\partial
x^{2}}+m^{2}\left(1-\frac{2}{\cosh^{2}(mx)}\right)\right]\eta_{n}(x)=\omega_{n}^{2}\eta_{n}(x),
\end{equation}
that only has one discrete mode, i.e., the zero mode of translation
\begin{eqnarray}\label{16}
    \eta_{0}(x)=\sqrt{\frac{m}{2}}\,\mbox{sech}(mx), \hskip3cm \omega_{0}^{2}=0.
\end{eqnarray}
Now we consider this system in the GUP framework where the modified
Hamiltonian takes the form
\begin{equation}\label{17}
H=p^{2}+ \frac{2\beta}{3}
p^{4}+m^{2}\left(1-\frac{2}{\cosh^{2}(mx)}\right).
\end{equation}
The extra term $\frac{2}{3}\beta p^{4}$ in the Hamiltonian results
in a positive shift in the zero mode energy as
\begin{equation}\label{18}
\Delta\omega_{0}^{2}=\frac{2\beta}{3}\langle\eta_0|p^{4}|\eta_0\rangle=\frac{14}{45}\beta
m^{4}.
\end{equation}
Moreover, in the Hartree approximation, the zero mode energy
correction is given by \cite{28}
\begin{equation}\label{19}
\Delta\tilde\omega_{0}^{2}=30^{-2/3}g^{4/3}.
\end{equation}
So the comparison between these two results shows that for
\begin{equation}
\beta>0.33\frac{g^{4/3}}{m^4}.
\end{equation}
the gravitational effects that modifies the ordinary uncertainty
principle are more dominant than the quantum fluctuations.

\section{Conclusions}
In this paper, we have considered a GUP framework that admits a
nonzero minimal position uncertainty. Due to the presence of this
minimal length a term proportional to $ p^{4}$ is added to the
Hamiltonians of all quantum mechanical systems. Using the
perturbation theory we have obtained the effects of this extra term
on the $\lambda\phi^{4}$ kink and the sine-Gordon soliton energy
spectrum. We have compared our results with the ones that obtained
for the quantum fluctuations in 1-loop resummed and Hartree
approximations. We showed that for the $\lambda\phi^{4}$ kink the
effects of GUP are more important than 1-loop resummed and Hartree
approximations for $\beta>1.32\lambda/m^{4}$. Also, for the
sine-Gordon case, the GUP energy correction is more dominant than
the Hartree approximation for $\beta>0.33(g^{1/3}/m)^{4}$.

\end{document}